\documentclass[prl,onecolumn,showpacs]{revtex4}
\usepackage{epsfig}
\usepackage{amsbsy}
\usepackage{amsmath}
\usepackage{graphicx}
\usepackage{latexsym}
\usepackage{amssymb}
\usepackage{color}

\begin{document}

\title{Identifying the closeness of eigenstates in quantum many-body systems}
\author{Haibin Li$^{1}$, Yang Yang$^{1}$, Pei Wang$^{1}$ and Xiaoguang Wang$^{2}$}

\affiliation{$^{1}$Department of Applied Physics, Zhejiang
University of Technology, Hangzhou 310023, China.\\
$^{2}$Zhejiang Institute of Modern Physics, Department of Physics,
Zhejiang University, Hangzhou 310027, China}

\date{\today}

\begin{abstract}
We propose a new quantity called modulus fidelity to measure the
closeness of two quantum pure states. Especially, we use it to
investigate the closeness of eigenstates of quantum many-body
systems. When the system is integrable, the modulus fidelity of
neighbor eigenstates displays a large fluctuation. But the modulus
fidelity is close to a constant when system becomes non-integrable
with fluctuation reduced drastically. Average modulus fidelity of
neighbor eigenstates increases with the increase of parameters
that destroy the integrability, which also indicates the
integrable-chaos transition. In non-integrable case, it is found
two eigenstates are closer to each other if their level spacing is
small. We also show that the closeness of eigenstates in
non-integrable domain is the underlying mechanism of
\emph{eigenstate thermalization hypothesis} (ETH) which explains
the thermalization in nonintegrable system we studied.
\end{abstract}


\maketitle

\section{Introduction}

Statistical mechanic has succeeded in describing quantum many-body
system. But how to understand thermodynamics from quantum
mechanics remains an unsolved problem in physics. We speak of
`thermalization' for an isolated quantum many-body system when it
reaches a steady state that can be described by an equilibrium
thermodynamic ensemble and the expectation values of observables
are time-independent at this steady state. Recently, the ground
breaking experimental progress in preparing and manipulating
ultracold atomic quantum gases~\cite{Exp} have made it possible to
obtain a system weakly coupled with the environment. These
approximately isolated systems are ideal test beds for the
investigation of relaxation and thermalization. It is found that
the relaxation to equilibrium surely takes place~\cite{Hoff} in
certain setups but fails to be approached in others~\cite{Kino}.
These experimental results stimulate intense theoretical effort to
find the condition under which thermalization of quantum system
can be approached.

Integrability is believed to be the intrinsic mechanism of the
failure of thermalization in one-dimensional boson
system~\cite{Rig3}. It is also found that the breaking of
integrability is related to the onset of thermalization. On the
other hand, quantum chaos will emerge in the breaking of
integrability when the perturbation is strong enough. Then, the
link between quantum chaos and thermalization seems to be
established. As well known, quantum chaos is used to address some
properties of quantum system whose classic counterparts are
chaotic. It is found that the properties of spectra, eigenstate
and dynamics of quantum system can indicate the emergence of
quantum chaos. Most of these studies focus on the statistical
properties~\cite{Guhr,Santos1,Santos2,Santos4,Santos3}. An
important description of energy level is level spacing
distribution $P(s)$, which is the probability distribution of the
distances between consecutive levels $s_i=E_{i+1}-E_i$, where
$s_i$ is normalized to unit as $\langle s \rangle =1$. At
integrability, $P(s)$ is a typically Poisson distribution
$P(s)=\displaystyle e^{-s}$ while at chaos the distribution takes
the Wigner-Dyson form $P(s)=(\pi/2)s \displaystyle e^{-\pi
s^2/4}$~\cite{Guhr, Haake}. Consider the eigenstate expanded in
certain basis, the distribution of coefficient was shown to be
localized or extended, which is related to the integrable-chaos
transition to some degree. In latter case, eigenstates are chaotic
wavefunction ~\cite{Casati,Bor}and they are similar because of
similar structure~\cite{Santos2,Santos4},which is a kind of
statistical similarity. However, despite intense studies, what is
the nature of the integrable-chaos transition is still unclear.

For an isolated quantum many-body system with Hamiltonian $H$, the
eigenstate thermalization hypothesis(ETH)~\cite{ETH} states that
thermalization occurs in individual eigenstates. That is to say,
let $\vert\Psi_i \rangle$ denote an eigenstate of the system with
eigenvalue $E_i$ and let $A$ denote a few-body observable, then
the equation
\begin{equation}
\langle\Psi_i\vert  A \vert\Psi_i\rangle=\langle  A
\rangle_{micro}(E_i)
\end{equation}
is satisfied, where the subscript \textit{micro} denotes the
microcanonical thermal average over the energy interval
$[E_i-\triangle E /2,E_i+\triangle E/2]$, where $\triangle E$ is
the energy interval. ETH has been verified numerically in a wide
variety of quantum many-body systems which are far from
integrability~\cite{Rig1, Rig2,Santos2,Santos4}. Recently, it was
shown~\cite{Rig4}that ETH is essentially equivalent to the basic
assumption of von Neumann's quantum ergodic
theorem(QET)~\cite{Neumann}

ETH which shows the closeness of expectation value of observable
on energy eigenstate motivates us to study a definite closeness of
energy eigenstates close in energy, other than statistical
similarity. The intrinsic and widely used tool to investigate the
closeness of two states,including pure state and mixed state, in
quantum information theory is fidelity $F$~\cite{Nielsen}. As
previously discussed, experiment setup can prepare isolated
quantum many-body system which is described by a single wave
function according to quantum mechanics. Therefor, in the
following, we just consider pure state. Assuming two pure states
denoted by $\vert \Phi \rangle$ and $\vert \Psi \rangle$, the
fidelity of these two states is defined as $F(\vert \Phi
\rangle,\vert \Psi \rangle)=|\langle \Phi \vert \Psi \rangle |$.
If these two states are identical, $\vert \Phi \rangle = \vert
\Psi \rangle $ or differ by a global phase as $\vert \Phi \rangle
= e^{i\theta}\vert \Psi \rangle $, then $F=1$. If they are
orthogonal, $F=0$. As all eigenstates of one quantum many-body
system are orthogonal, the fidelity of each pair of eigenstates
are all zero no matter what this system is, integrable or
nonintegrable, then we cannot use fidelity to define and study the
closeness of eigenstates of quantum many-body system in this
sense. To study the substantial properties of quantum many-body
system,for example, the transition from integrable to chaos, we
should build a new understanding of closeness of quantum stats.

This article is organized as follows. In Sec.
\uppercase\expandafter{\romannumeral2}, we introduce a new
measurement called modulus fidelity. In Sec.
\uppercase\expandafter{\romannumeral3},we use modulus fidelity to
investigate the transition from integrable to chaos in
one-dimensional quantum many-body system. In
Sec.\uppercase\expandafter{\romannumeral4},based on numerical
results, we provide an understanding of the origin of ETH.
Finally, we summarize our results in
\uppercase\expandafter{\romannumeral5}.

\section{The definition of modulus fidelity}
In this paper, we propose a new measurement to define and identify
the closeness of two quantum states which are the eigenstates of
an isolated quantum many-body system with Hamiltonian $ H $. We
still denote them by $\vert \Phi \rangle$ and $\vert \Psi \rangle$
. Let $\vert n_i \rangle $ denotes a complete orthogonal set in
Hilbert space of this quantum system with dimension $L$. This
orthogonal basis is arbitrary in Hilbert space and can be assumed
as the eigenstates of an observable $A$ with eigenvalue $a_i$.
Then $\vert \Phi \rangle $ and $\vert \Psi \rangle $ can be
expanded as $\vert \Phi \rangle =\sum_{i=1}^L c_i \vert n_i
\rangle, $ and $\vert \Psi \rangle =\sum_{j=1}^L d_j \vert n_j
\rangle $, where $c_i$($d_j$) is the expansion coefficient. The
fidelity of these two state is $F=\sum_{i=1}^L c_i^*d_i$. If we
measure $A$ on the state $\vert \Psi \rangle$, then the
probability of obtaining eigenvalue $a_i$ is $\vert c_i \vert ^2$.
After repeating measurement, one can get all eigenvalues with a
sequence of probability $\{\vert c_1 \vert ^2,\vert
c_2\vert^2,\cdot \cdot \cdot \cdot \vert c_L \vert^2\}$. Doing
same measurement on state $\vert \Phi \rangle$, we can get another
sequence $\{\vert d_1 \vert ^2,\vert d_2 \vert ^2, \cdot \cdot
\cdot \cdot\vert d_L \vert^2\}$. To identify the closeness of two
state by comparing these two sequence, we use a measure in the
form of fidelity as
\begin{equation}
F_m=\sum_{i=1}^L \vert c_i \vert \vert d_i \vert
\end{equation}
where the modulus operation is to make the value of $F_m$ in
$[1,0]$ as well as fidelity and we call $F_m$ modulus fidelity.
Such measure is similar with the one used in
ref.~\cite{Fyod,Santos3} where the product of square of modulus of
expansion coefficient were summed.

To better understand this new measure, we construct a new kind of
vectors in the same Hilbert space $\vert n_i \rangle $ by setting
its coefficients as the modulus of the coefficients of a vector
like $\vert \Phi \rangle $ . Then the new vectors built on $\vert
\Phi \rangle $ and $\vert \Psi \rangle $, denoting by $\vert
\widetilde{\Phi} \rangle$ and $\vert \widetilde{\Psi} \rangle $,
are defined as

\begin{eqnarray}
\vert \widetilde{\Phi} \rangle =\sum_i^L \vert c_i \vert \vert n_i \rangle, \nonumber \\
\vert \widetilde{\Psi} \rangle  =\sum_j^L \vert d_j \vert \vert
n_j \rangle.
\end{eqnarray}
then  $F_m$ of two pure states $\vert \Phi \rangle $ and $\vert
\Psi \rangle $ is the fidelity of $\vert \widetilde{\Phi} \rangle$
and $\vert \widetilde{\Psi} \rangle $ as

\begin{equation}
F_m(\vert \Phi \rangle,\vert\Psi \rangle)=|\langle
\widetilde{\Phi} \vert \widetilde{\Psi} \rangle|=\langle
\widetilde{\Phi} \vert \widetilde{\Psi} \rangle .
\end{equation}
In last equality, the modulus operation vanishes due to the
definition of new state (3). Note that $F_m$ has the same
definition as fidelity, but it is defined in new constrained
space, describing the overlap of two quantum pure states in terms
of modulus of coefficient. Modulus fidelity has some different
properties from fidelity, for example, it is dependent on the
choice of basis, while fidelity is independent. Such replacing
coefficients by their modulus was also proposed recently in
ref.\cite{Fish} to study its effect on entanglement entropy.

\section{Numerical results}

In the following, we use modulus fidelity $F_m$ to study the
closeness of eigenstates of a quantum many-body system in both
integrable and nonintegrable domain. We consider one-dimensional
hard-core boson model(HCB) with dimensionless Hamiltonian

\begin{eqnarray}
H=\sum_{i=1}^N \{-t(\hat{b}_i^\dagger
b_{i+1}+H.c.)-t'(\hat{b}_i^\dagger b_{i+2}+H.c.)\nonumber
+V\hat{n}_i \hat{n}_{i+1}+V'\hat{n}_i \hat{n}_{i+2}\},
\end{eqnarray}
where $t$ and $t'$ are the nearest-neighbor hopping  and the
next-nearest-neighbor hopping, $V$ and $V'$ are the
nearest-neighbor and the next-nearest-neighbor interaction
respectively. Throughout this paper, $t$ and $V$ are set to be
unit, $t=V=1$, and $t'$ and $V'$ are set to be equal, $t'=V'$.
$\hat{n}_i=\hat{b}_i^\dagger \hat{b}_i$ is density operator. As
well known, this model is integrable when $t'=V'=0.0 $ but
nonintegrable when $t'=V'\neq 0.0$. It has been verified that the
thermalization can be achieved in this system and the underlying
mechanism is ETH when system is far from
integrable~\cite{Rig1,Rig2,Santos2}.

\begin{figure}
\includegraphics[width=0.6\textwidth,height=0.5\textwidth]{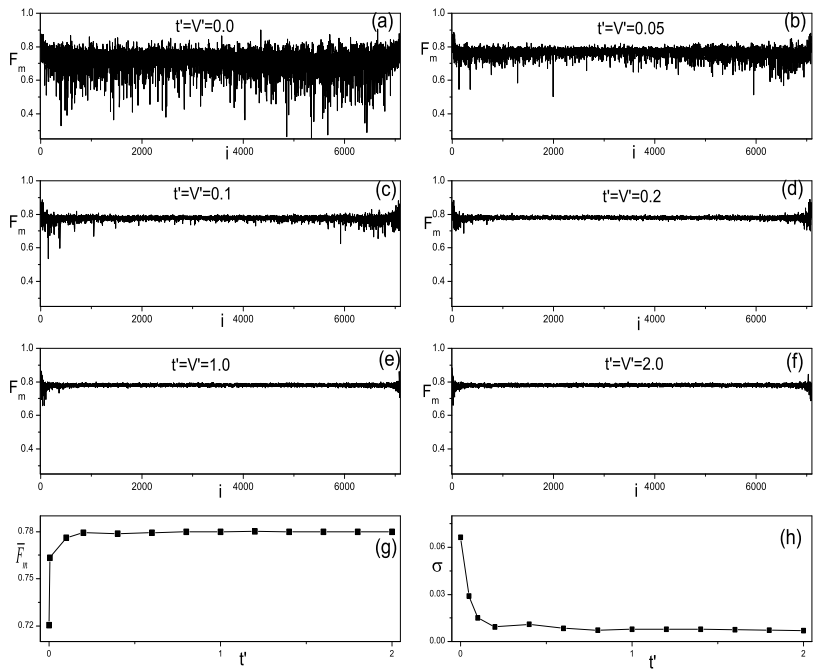}
\caption{The modulus fidelity $F_m(i,i-1)$ of $\textit{ith}$
eigenstates and $\textit{(i-1)th} $ eigenstate in 1-D hard-core
model. The data are obtained in the condition that 6 bosons on 25
lattice and the next-nearest-neighbor parameters are changed from
(a) integrable case to nonintegrable case (b)-(f). Average modulus
fidelity (g) and the standard deviation of modulus fidelity (h) of
neighbor eigenstates in present model as a function of the
next-nearest-neighbor hopping parameters $t'=V'$.}

\end{figure}

We use full exact diagonalization method to calculate all
eigenstates and eigenvalues of present system. We study the
lattice up to 25 sites and 6 hard-core bosons. Under period
boundary condition, the system preserves translational symmetry,
by which the Hilbert space of Hamiltonian can be decomposed into
different independent subspace with different total momentum $k$.
We can diagonalize each subspace. As discussed above, modulus
fidelity is dependent on the choice of basis. That is to say, for
two eigenstates we consider, the values of modulus fidelity is
different in different basis. However, as the number of
eigenstates of many-body quantum system is large, we can focus on
the statistical properties of modulus fidelity of eigenstates. We
investigate the modulus fidelity in two different basis, site
basis(Fock state) and k basis(Bloch state). These two bases were
used to study the localization of a single eigenstate of quantum
many-body system in integrable-chaos transition.

In our simulation, we found that the statistical properties of
modulus fidelity are the same in these two kinds of bases.
Throughout this paper, we just illustrate the results of modulus
fidelity obtained in site basis. The eigenstates in illustration
are obtained in subspace with momentum $k=1$ rather than $k=0$ to
avoid a parity symmetry. Then they are transformed into to space
of site basis.

\begin{figure}
\includegraphics[width=0.45\textwidth]{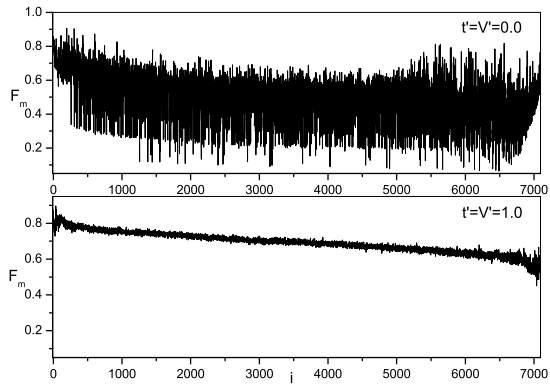}
\caption{The modulus fidelity of excited stated with the ground
state. (a) $t'=V'=0.0$ and (b)$t'=V'=1.0$.}
\end{figure}

We first calculate $F_m$ of neighbor eigenstates in subspace at
different values of the next-nearest neighbor hopping and
interaction which determines the breaking of integrability in the
system. The numerical results are plotted in Fig.1. Note that when
$t'=V'=0$, i.e., the system is integrable, as shown in Fig.1(a),
each modulus fidelity $F_m$ of the pair of neighbor eigenstates is
non-zero, and most of them are larger than 0.5, which means, on
the whole, that in integrable case eigenstates are close to their
neighbor ones in term of modulus fidelity. The most significant
feature is that they have large fluctuation. $F_m$ distribute
along the eigenvalues like random number in an interval with large
fluctuation. When $t'$ and $V'$ are non-zero, the system becomes
nonintegrable. As is well known, the level spacing distribution
changes from Poisson to Wigner-Dyson form, indicating a transition
from integrable to chaos. In Fig.1 (b-f), we found that the values
of modulus fidelity between most neighbor eigenstates are also
larger than 0.5 and vary round a constant as well as the case of
integrability. However, there is a dramatic change appearing that
the fluctuation of modulus fidelity is reduced with the increase
of $t'$ and $V'$. When $t'$ and $V$ is large enough, for example,
$t'=V'=2.0$ in Fig.1(f), as a function of index of eigenvalues,
$F_m$ is no longer like random number, but tends to be a smooth
constant. That is to say, when the system comes into chaos, most
of eigenstates of system becomes close to their neighbor one in
term of modulus fidelity and the values of the modulus fidelity
also become close so that eigenstate are close to each other.

We also calculate the average modulus fidelity of all modulus
fidelity between pair neighbor eigenstates,
$\overline{F}_m=\frac{1}{L-1}\sum _{i=1}^{L-1}F_m(i)$, where $L$
is the size of subspace, and the standard deviation. The results
are also plotted in Fig 1. We can see obvious changes in both
measures. When $t'$ and $V'$ increase from zero,i.e.,
integrability point, average modulus fidelity increases quickly.
In Fig.1(a-f), we have found all eigenstates become close to each
other when system is far from integrability. Here, we find that
the degree of closeness of eigenstates will also increase in this
process. After $t'$ an $V'$ reach a certain value, average modulus
fidelity will not increase so quickly and almost tends to be
saturated. For the parameters of model we investigated here, this
value is about 0.25. This scenario can also be seen in the results
of standard deviation of modulus fidelity. It decreases very
quickly when $t'$ and $V'$ increase from zero, which indicates the
reduction of fluctuation seen in Fig.1(a-f). The reduction of
fluctuation was also found in ref. \cite{Santos3} by using a
similar measure as discussed above. But except for this result,
the closeness are measured differently. In ref. \cite{Santos3},
maximal value of overlap(closeness) decrease when system becomes
nonintegrable, which is contrary to our result. When $t'$ and $V'$
are large enough and system comes into chaos, the fluctuation is
inhibited to be small and decreases slowly as confirmed in
Fig.1(h). The values of $t'$ and $V'$ at which transition occurs
are the same as that of the average modulus fidelity. The
existence of such critical values has been
found~\cite{Rig2,Santos2,Rig3} and they will decrease with the
increase of size of system. So it is believed that in the
thermodynamic limit an infinitesimal integrability breaking would
lead to chaos.

\begin{figure}
\includegraphics[width=0.45\textwidth]{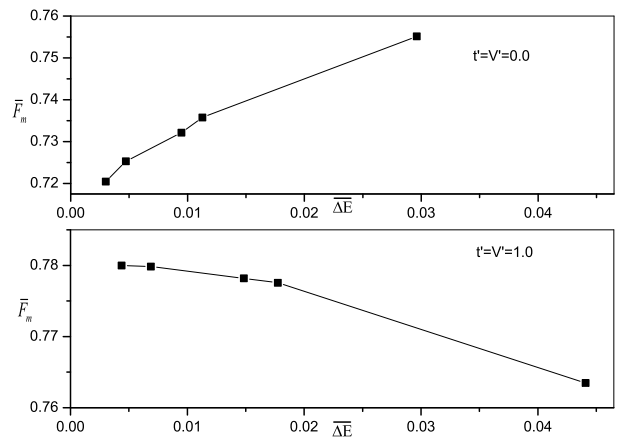}
\caption{Average modulus fidelity versus average level
spacing(energy density) for 1-D hard-core boson system with
(a)$t'=V'=0.0$ and (b)$t'=V'=1.0$. Average level spacings
$\overline{\Delta E}$ are obtained by $(E_{max}-E_{min})/N $ where
$E_{max}$ and $E_{min}$ are the two ends of energy spectrum.  }.
\end{figure}
As shown previously , when system is in integrable and
nonintegrable cases, the statistical properties of modulus
fidelity of neighbor eigenstates behave differently. Next, we
provide further evidence for this statement. We choose ground
state as a reference state and calculate modulus fidelity of
excited eigenstates with respect to it. We plot some results of
modulus fidelity as a function of the index of eigenstate in
Fig.2. Both in integrable case and nonintegrable case, the overall
trend of modulus fidelity of excited state with respect to ground
state decreases when the excited state is far from ground state in
energy. However, the local details in two case are different. We
also notice a reduction of fluctuation appearing in such modulus
fidelity as well as modulus fidelity of neighbor eigenstates shown
in Fig.1. In integrable case, there is also a large fluctuation
presenting along the function of modulus fidelity. That is to say,
there is no clear relation between closeness and level spacing.
When system is nonintegrable, the fluctuation is reduced sharply
and modulus fidelity of excited state with respect to ground state
becomes a smooth function of index of eigenstate. That is to say,
the closeness of two states and level spacing are in inverse
proportion,i.e,the small level spacing is, the more close two
states become. The eigenvalues of one Hamiltonian have a nature
order, varying from minimum to maximum. Due the results show in
Fig.1 and Fig.2, we find that at chaos such order of eigenvalues
has its counterpart in eigenstates, which is described by the
modulus fidelity. However, this relationship is destroyed in
integrable case, by which we can also distinguish the integrable
and chaos domain in a quantum many-body system.

\begin{figure}
\includegraphics[width=0.45\textwidth]{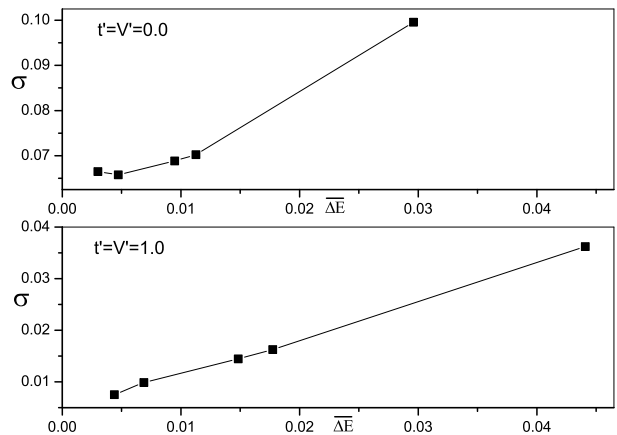}
\caption{Standard deviation versus average level spacing(energy
density) for 1-D hard-core boson system with (a)$t'=V'=0.0$ and
(b)$t'=V'=1.0$. Average level spacings $\overline{\Delta E}$ are
obtained by $(E_{max}-E_{min})/N $ where $E_{max}$ and $E_{min}$
are the two ends of energy spectrum. }.
\end{figure}

Next, we investigate the relation between the closeness of
eigenstates and level spacing by changing the system size. As well
known , in the thermodynamic limit, the level spacing of quantum
many-body is expected to be zero. For a finite system, level
spacings between each pair of neighbor eigenstates are not the
same and obey a distribution. To compare the closeness at
different system size, we calculate average modulus fidelity of
neighbor eigenstates. Accordingly, level spacings are also
averaged. In Fig.3, we plot numerical results of system in both
integrable and nonintegrable cases. When system is integrable,
average modulus fidelity decrease with the decrease of energy
level, which means the close of energy level lead to the
separation of corresponding eigenstates. But in nonintegrable
case, average modulus fidelity increases with the decrease of
level spacing, showing that when eigenstates become close in
energy, they will be close in term of modulus fidelity as well as
in Fig.2. The standard derivation of modulus fidelity of neighbor
eigenstates in different system size are plotted in Fig.4. When
system is nonintegrable, standard derivation decreases with the
decrease of level spacing, i.e., the increase of system size. That
is to say, when system size becomes large, the closeness of
neighbor eigenstates tends to be the smooth function of energy.
When system is integrable, standard derivation also decreases as
the the system size increase for most data, except for the one
when system size takes maximal value in our study. In the future
work, a large scale simulation should be done to find whether it
is a error or a intrinsic character of integrable system.

\section{The underlying mechanism of Eigenstate Thermalization Hypothesis}

We have identify the integrable-chaos transition in quantum many
body system which is related to the emergence of thermalization.
As ETH states, when system is at chaos, the expectation value of
the observable $A$ on energy eigenstate $\vert  \Phi_i \rangle$,
$\langle \Phi_i \vert A \vert \Phi_i \rangle$, is a smooth
function of eigenvalues $E_i$. One can expand eigenstate $\vert
\Phi_i \rangle$ by another orthogonal complete basis which could
be constructed by the eigenstates of observable $A$. If $A$
commutes with Hamiltonian, its eigenstates are the same as that of
Hamiltonian, so $\vert \Phi_i \rangle$ is expanded on itself. In
this case, $F_m$ is zero for any pairs of eigenstates, which is
trivial for studying the closeness of eigenstate as well as
fidelity $F$. We focus on the case that observable $A$ does't
commuted with Hamiltonian. For the quantum system we considered,
assume an eigenstate of observable $A$ is $\vert \alpha_l \rangle
$ with eigenvalues $a_l$, i.e., $A \vert \alpha_l \rangle=a_l
\vert \alpha_l \rangle$. Let $\vert \Phi_i \rangle$ and $\vert
\Phi_{i-1} \rangle$ denote two eigenstates of Hamiltonian close in
energy, then they can be expanded by the eigenstates of observable
$A$ as $\vert \Phi_i \rangle =\sum_{l} \beta_{i,l} \vert \alpha_l
\rangle, \vert \Phi_{i-1} \rangle =\sum_{l} \beta_{i-1,l} \vert
\alpha_l \rangle $ where $\beta_{n,l}$ is coefficient. Then the
expectation value of $A $ on these two eigenstates are

\begin{eqnarray}
\langle \Phi_i \vert A \vert \Phi_i \rangle=\sum_{l} \vert
\beta_{i,l} \vert^2 a_l.    \nonumber\\
\langle \Phi_{i-1} \vert A \vert \Phi_{i-1} \rangle=\sum_{l} \vert
\beta_{i-1,l} \vert^2 a_l.
\end{eqnarray}
where $\vert \beta_{i,l} \vert ^2 =\beta_{i,l}^*\beta_{i,l} $  is
the eigenstate occupation numbers which is the probability of
obtaining the eigenvalue $a_l$ in a measurement of $A$ on energy
eigenstate $\vert \Phi_i \rangle$. In nonintegrable case, due to
Eq.(1) of ETH, $A(E)$ equals to microcanonical average, which
means
\begin{equation}
\langle \Phi_i \vert A \vert \Phi_i \rangle \approx \langle
\Phi_{i-1} \vert A \vert \Phi_{i-1} \rangle.
\end{equation}
Due to Eq. (5), one can find the valid of Eq. (6) is contributed
by two properties, the eigenvalues of $A$ and eigenstate
occupation number. As the former is fixed for a quantum system,
then there should be two possible scenario for eigenstate
occupation number($\vert \beta_{n,l} \vert ^2)$ that they
fluctuate between eigenstate close in energy or not. Now, our
result that in nonintegrable system we studied the modulus
fidelity has larger value and keeps almost constant suggests the
later that for two eigenstates of Hamiltonian close in energy, the
eigenstate occupation number of them are also close as
\begin{equation}
\vert \beta_{i,l} \vert ^2\ \approx \vert \beta_{i-1,l} \vert ^2.
\end{equation}
Then the Eq. (6) will be satisfied. In this case, the modulus
fidelity between each pair of neighbor eigenstate will tend to be
unit and equal to each other, showing a smooth function versus
eigenvalues. On the contrary, if the expansion coefficient $\vert
\beta_{n,l} \vert ^2)$ is fluctuating between eigenstates close in
energy, i.e., Eq. (7) is violated, the modulus fidelity between
pair neighbor eigenstates will have a fluctuation as shown in
Fig.1 to Fig.3. In ref.\cite{Rig4}, the basis of ETH and QET is
assumed that the overlap between an energy eigenstate and an
eigenstate of an observable A is exponential small and close to
$\frac{1}{D}$, where $D$ is the dimension of Hilbert space of
system. It can be found here that this assumption is an extreme
point in our result that will also make modulus fidelity to be
unit.

\section{Conclusion}
In conclusion, we found that the closeness of eigenstates of
Hamiltonian of quantum many-body system can be studies as an
indicator for the transition from integrability to chaos. For this
purpose, we proposed a new measure, modulus fidelity, which is
defined by replacing the expansion coefficients of one quantum
state in a basis by their modulus. The full exact diagonalization
of one-dimensional hard-core boson model showed that in integrable
and nonintegrable case, the modulus fidelity of neighbor
eigenstates have different properties, which indicates the
integrable-chaos transition. The reduction of fluctuation of
modulus fidelity and the increase of their average value in this
transition show the eigenstates of quantum many body system  tend
to be close to each other, which show a deeper uniformization of
eigenstates than statistical similarity. We also studied the
finite effect on such transition. Furthermore, our analysis show
the closeness of eigenstate guarantees the valid of ETH in
nonitegrable system we studied and suggest a general understanding
of the underlying mechanism of ETH and QET.


\begin{thebibliography}{99}

\bibitem{Exp}Greiner M, Mandel O, Hansch T W and Bloch I 2002 Nature {\bf
419 51};  Kinoshita T, Wenger T and Weiss D S 2006 Nature {\bf 440
900};  Sadler L E, Higbie J M, Leslie S R, Vengalattore M and
Stamper-Kurn D M 2006 Nature {\bf 443 312}; Ritter S, Ottl A,
Donner T, Bourdel T, K$\ddot{o}$hl M and Esslinger T 2007 Phys.
Rev. Lett. {\bf 98 090402};  Trotzky S, Chen Y -a, Flesch A,
McCulloch I P, Schollw$\ddot{o}$ck U, Eisert J and Bloch I 2012
Nature Phys. {\bf 8 1}


\bibitem{Hoff}Hofferberth S, Lesanovsky I, Fischer B, Schumm T and
Schmiedmayer J 2007 Nature {\bf 449 324}

\bibitem{Kino}Kinoshita T, Wenger T and Weiss D S 2006 Nature {\bf 440 900}

\bibitem{Rig3} Rigol M, Dunjko V, Yurovsky V and Olshanii M 2007 Phys. Rev. Lett. {\bf 98 050405};
Cazalilla M A 2006 Phys. Rev. Lett. {\bf 97 156403};  Barthel T
and Schollw\"ock U 2008 Phys. Rev. Lett.{\bf 100 100601}; Kollar M
and Eckstein M 2008 Phys. Rev. A {\bf 78 013626};  Iucci A and
Cazalilla M A 2009 Phys. Rev. A.{\bf 80 063619}


\bibitem{Guhr} Guhr T, M\"uller-Groeling A and
Weidenm\"uller H A 1998 Phys. Rep. {\bf 299 189}

\bibitem {Santos1}Santos L F, Borgonovi F and Izrailev F M 2012
Phys. Rev. Lett. {\bf 108 094102}

\bibitem{Santos2}Santos L F and Rigol M 2010 Phys. Rev. E. {\bf
81 036206}

\bibitem{Santos4}Santos L F and Rigol M 2010 Phys. Rev.
E {\bf 82 031130}

\bibitem{Santos3}Santos L F, Borgonovi F and Izrailev F M 2012 Phys. Rev. E.{\bf
85 036209}



\bibitem{Haake} F. Haake 1991 Quantum Signatures of Chaos (Spinger-Verlag, Belin)


\bibitem{Casati}Casati G, Chirikov B V, Guarneri I and Izrailev F M 1993 Phys. Rew. E. {\bf
48 R1613}; 1996 Phys. Lett. A {\bf 223 430}

\bibitem{Bor} Borgonovi F, Guarneri I and Izrailev F M 1998 Phys.
Rev. E {\bf 57 5291}; Luna-Acosta G A, M¨¦ndez-Berm¨²dez J A and
Izrailev F M 2000 Phys. Lett. A, {\bf 274 192}; Benet L, Izrailev
F M, Seligman T H and Su\' arez-Moreno A 2000 Phys. Lett. A, {\bf
277 87}

\bibitem{ETH} Deutsch J M 1991 Phys. Rev. A {\bf 43 2046};  Srednicki M 1994 Phys. Rev. E {\bf 50
888}

\bibitem{Rig1} Rigol M, Dunjko V and  Olshanii M 2008 Nature {\bf 52 854}

\bibitem{Rig2} Rigol M 2009 Phys. Rev. Lett. {\bf 103 100403};
 Rigol M 2009 Phys. Rev. A {\bf 80 053607}; Rigol M
and Santos L F 2010 Phys. Rev. A {\bf 82 011604}

\bibitem{Rig4} Rigol M and Srednicki M 2012 Phys. Rev. Lett. {\bf 108 110601}

\bibitem{Neumann}Neumann J von  2010 European Phys. J. H {\bf 35 201}

\bibitem{Nielsen} Nielsen M A and Chuang L I 2000 Quantum Computation and
Quantum Information (Cambridge University Press, Cambridge,
England)

\bibitem{Fish} T. Grover and P. A. Fisher,  arXiv:1412.3534


\end{thebibliography}
\end{document}